\documentstyle{./aa}
\begin{document}

   \title{A Deep H$\alpha$ Survey of Galaxies in the Two Nearby Clusters
Abell~1367 and Coma}

   \subtitle{The H$\alpha$ Luminosity Functions}

   \author{J. Iglesias-P\'{a}ramo
          \inst{1}
          \and
A. Boselli
	\inst{1}
	\and
L. Cortese
	\inst{2}	
	\and
J.M. V\'{\i}lchez
	\inst{3}
	\and	
G. Gavazzi
	\inst{2}
          }

   \offprints{J. Iglesias-P\'{a}ramo}

   \institute{
Laboratoire d'Astrophysique de Marseille, Traverse du Siphon - 
Les Trois Lucs, 13376 Marseille, France\\
              \email{jorge.iglesias@astrsp-mrs.fr,alessandro.boselli@astrsp-mrs.fr}
         \and
	 Universit\`a degli Studi di Milano - Bicocca, P.zza delle scienze 3,
20126 Milano, Italy.\\
	\email{giuseppe.gavazzi@mib.infn.it,luca.cortese@mib.infn.it}
	\and
             Instituto de Astrof\'{\i}sica de Andaluc\'{\i}a (CSIC), Granada, Spain\\
             \email{jvm@iaa.es}
            }

   \date{Received ......; accepted .........}

   \abstract{
We present a deep wide field H$\alpha$ imaging survey of the
central regions of the two nearby clusters of galaxies Coma and Abell~1367, taken with the
WFC at the INT~2.5m telescope. 
We determine for the first time the Schechter parameters of the H$\alpha$
luminosity function (LF) of cluster galaxies.
The H$\alpha$ LFs of Abell~1367 and Coma are compared with each other and with that of
Virgo, estimated using the $B$ band LF by Sandage et al. (1985) and a
$L$(H$\alpha$)~$vs$~$M_{B}$ relation.
Typical parameters of $\phi^{*} \approx 10^{0.00\pm0.07}$~Mpc$^{-3}$, $L^{*} \approx
10^{41.25\pm0.05}$~erg~sec$^{-1}$ and $\alpha \approx -0.70\pm0.10$ are found for the three
clusters. The best fitting parameters of the cluster LFs differ from
those found for field galaxies, showing flatter slopes and lower scaling
luminosities $L^{*}$. 
Since, however, our H$\alpha$ survey is significantly deeper than
those of field galaxies, this result must be confirmed on similarly deep
measurements of field galaxies.
By computing the total SFR per unit volume of cluster galaxies, and taking into
account the cluster density in the local Universe, we estimate that the
contribution of clusters like Coma and Abell~1367 is approximately 0.25\%
of the SFR per unit volume of the local Universe.
   }

   \maketitle

\section{Introduction}

The strong morphology segregation observed in rich clusters of galaxies
(Dressler, 1980) testifies the fundamental role played by the environment
on the evolution of galaxies. Which physical mechanisms are responsible
for such transformations is however still matter of debate.
Several processes might alter the evolution of cluster galaxies. Some
of them refer to the interaction of the galaxies with the intracluster
medium (Gunn \& Gott, 1972) and others account for the effects of
gravitational interactions produced by the gravitational potential of 
the cluster (Merritt, 1983)
or by galaxy-galaxy interactions (Moore et al. 1996, 1998, 1999).
All these mechanisms can produce strong perturbations in the galaxy 
morphology with the formation
of tidal tails, dynamical disturbances which appear as asymmetries in 
the rotation curves
(Dale et al. 2001) and significant gas removal (Giovanelli \& Haynes 
1985; Valluri \& Jog 1990).

Some of these processes are expected to produce changes in the star 
formation rates
of galaxies in clusters. 
Several studies have addressed the issue of the influence of the cluster
environment on the SFR of disk galaxies, however no agreement has been 
established so far:
whereas some authors proposed similar or even enhanced star formation in
cluster spirals than in the field (Donas et al. 1990, 1995; Moss \& 
Whittle 1993,
Gavazzi \& Contursi 1994; Moss et al. 1998;
Gavazzi et al. 1998; Moss \& Whittle 2000), some others claim quenched 
SFRs in
cluster spirals (Kennicutt 1983; Balogh et al. 1998; Hashimoto et al. 1998).
This discrepancy could arise from non-uniformity of the adopted methods (UV
vs. H$\alpha$ vs. [O{\sc ii}] data) or from real differences in the
studied clusters (Virgo, Coma, Abell~1367, clusters from Las Campanas 
Redshift
Survey, clusters at $z > 0.18$).\\

In particular, an enhanced fraction of spirals with
circumnuclear H$\alpha$ emission was found in the highest density
regions of some nearby clusters (Moss et al. 1998; Moss \& Whittle 2000),
whereas no such difference was found for galaxies with diffuse emission.
The compact H$\alpha$ emission seems associated
with ongoing interactions of galaxies, but numerical simulations by Bekki
(1999) showed that mergers between clusters and subclusters might produce
central starbursts in cluster spirals.

Existing studies of the H$\alpha$ properties of
galaxies in clusters suffer from various biases: the photoelectric data by
Kennicutt et al. (1984) and Gavazzi et al. (1991, 1998) are based on 
samples of
galaxies selected on the basis of their optical properties, independent 
of their
H$\alpha$ properties. On the other hand, the objective-prism surveys by 
Moss et
al. (1988, 1998) and Moss \& Whittle (2000) are H$\alpha$ selected 
but they are too shallow to allow a determination of the H$\alpha$
luminosity function as deep as desired.\\
With the aim of obtaining a reliable determination of the current SFR in 
nearby clusters
of galaxies and to study the spatial distribution of the star formation 
regions,
we undertook a deep imaging survey of a one degree $\times$ one degree 
area of
the Coma and Abell~1367 clusters.\\
Our work provides the first deep and complete study of galaxies in 
clusters based on their H$\alpha$
emission properties.\\

This paper is arranged as follows: Section~2 contains a description of 
the observations, of
the data reduction and the detection procedures.
The H$\alpha$ data are presented in Section~3. The H$\alpha$
luminosity function and a brief discussion on the contribution of both 
clusters
to the local star formation rate density are presented in
Section~4. Conclusions are presented in Section~5.
Comments on the most interesting objects as well as the H$\alpha$ images 
of the
detected galaxies are given in the Appendix.

   \begin{figure}[t]
   \centering
      \caption{
Transmitance of the filters used for the observations.
}
         \label{filters}
   \end{figure}

\section{Observations and Data Reduction}

   \begin{table}[t]
      \caption[]{Journal of the Observations}
         \label{log}
     $$ 
         \begin{tabular}{lllll}
            \hline
            \noalign{\smallskip}
Field & R.A. & Dec. & Exp. & Filter \\
      & (J2000)     & (J2000)     & sec     &        \\
            \noalign{\smallskip}
            \hline
            \noalign{\smallskip}
\multicolumn{5}{c}{26th April 2000}\\
            \hline
            \noalign{\smallskip}
Coma~1 & 12:59:24.75 & $+$27:58:49.89 & $3\times 1200$ & [S{\sc ii}] \\
& & & $3\times 300$ & $r'$ \\
Coma~2 & 13:01:24.45 & $+$27:58:52.12 & $1200$ & [S{\sc ii}] \\
& & & $300$ & $r'$ \\
\hline
            \noalign{\smallskip}
\multicolumn{5}{c}{28th April 2000}\\
            \hline
            \noalign{\smallskip}
Coma~3 & 13:01:24.26 & $+$28:28:52.12 & $3\times
1200$ & [S{\sc ii}] \\
& & & $3\times 300$ & $r'$ \\
Coma~4 & 12:59:24.57 & $+$28:58:49.89 & $3\times
1200$ & [S{\sc ii}] \\
& & & $3\times 300$ & $r'$ \\
\hline
            \noalign{\smallskip}
\multicolumn{5}{c}{26th April 2000}\\
            \hline
            \noalign{\smallskip}
A1367~1 & 11:41:35.83 & $+$19:58:21.44 & $3\times
1200$ & [S{\sc ii}] \\
& & & $3\times 300$ & $r'$ \\
A1367~2 & 11:43:35.61 & $+$19:58:20.73 & $3\times
1200$ & [S{\sc ii}] \\
& & & $3\times 300$ & $r'$ \\
\hline
            \noalign{\smallskip}
\multicolumn{5}{c}{28th April 2000}\\
            \hline
            \noalign{\smallskip}
A1367~3 & 11:45:35.40 & $+$19:58:20.10 & $3\times
1200$ & [S{\sc ii}] \\
& & & $3\times 300$ & $r'$ \\
A1367~4 & 11:43:35.56 & $+$19:28:20.73 & $3\times
1200$ & [S{\sc ii}] \\
& & & $3\times 300$ & $r'$ \\
            \hline
            \noalign{\smallskip}
         \end{tabular}
     $$ 
   \end{table}

The observations were carried out with the Wide Field Camera (WFC) at
the Prime Focus at the INT~2.5m telescope located at Observatorio de El Roque de
los Muchachos (La Palma), on April 26th and 28th 2000, under photometric
conditions. The average seeing ranged from
1.5 to 2~arcsecs during both nights. Given the mean velocity of the galaxies in 
the two clusters under study, $6555~\pm~684$~km~sec$^{-1}$ and
$6990~\pm~821$~km~sec$^{-1}$ for Abell~1367 and Coma respectively\footnote{for
Abell~1367, we use the average of the redshifts reported in this paper.} (Fadda et
al. 1996), 
the narrow-band [S{\sc ii}] filter ($\lambda_{0} = 6725$\AA, $\delta \lambda
\approx 80$\AA) was used to isolate the
H$\alpha$ line and the $r'$ Sloan-Gunn broad-band filter ($\lambda_{0} =
6240$\AA, $\delta \lambda \approx 1347$\AA) to recover the continuum. 
Figure~\ref{filters} shows the transmitance profiles of both
filters. Given the width of the [S{\sc ii}] filter, the [N{\sc
ii}]$\lambda\lambda$6548,6584\AA\ lines are included in the high transmitance
pass-band of this filter, so in what follows we will refer to the combined
H$\alpha$~$+$~[N{\sc ii}] flux and equivalent width, as H$\alpha$ flux and
equivalent width respectively.

The WFC is composed by a science array of four thinned AR
coated EEV 4K$\times$2K devices, plus a fifth acting as
autoguider. The pixel scale is 0.333
arcsec~pixel$^{-1}$, giving a total field of view of about $34\times 34$~
arcmin$^{2}$. Given the arrangement of the detectors, a squared area
of about $11 \times 11$~arcmin$^{2}$ is lost at the top right corner of the
field. The top left corner of detector \#3 is also lost because of filter vignetting.

   \begin{figure*}
   \centering
      \caption{
Projected positions of our exposures in the Abell~1367 (left plot) and Coma
(right plot) clusters. Superimposed contours correspond to the X-ray maps from
Donnelly et al. (1998) and White et
al. (1993) respectively. Filled dots represent the
galaxies showing H$\alpha$ emission.
}
         \label{map}
   \end{figure*}

Four fields near the center of each cluster were
observed. Three different exposures, slightly dithered to remove
cosmic rays, were obtained for each position
in each filter, except for the second exposure of the Coma
cluster where 
only one exposure per filter was obtained. Figure~\ref{map} shows
our surveyed area.
Our observations cover mainly the North-East region of the Coma cluster as
described by Colless \& Dunn (1996), coinciding with the central part of the
Godwin catalog of the Coma cluster (Godwin et al. 1983). 
One of our fields of Abell~1367 (number 1 in Figure~\ref{map}) is not covered by the
Godwin catalog (Godwin \& Peach 1982). For comparison the X-ray contour maps of
the two clusters (White et al. (1993) for Coma, and Donnelly et al. (1998) for
Abell~1367) are plotted in the figure. 
The galaxies detected in H$\alpha$ are marked  with filled dots.
The diary of the observations is presented in Table~\ref{log}.

The data reduction was carried out using standard tools in the
IRAF\footnote{Image Reduction and Analysis Facility, written and supported at
the National Optical Astronomy Observatory} environment. 
The astrometric solution was found with the USNO\footnote{United States Naval
Observatory} catalog of stars. The accuracy of this solution was found to be
better than 3~arcsecs throughout the frames.
Several exposures of standard spectrophotometric stars were taken
during both nights. 
The chip-to-chip differential responses were derived by direct comparison of the
photometry measured for the objects, non-saturated stars and galaxies, present
in the overlapping regions.
Zero-points and extinction coefficients were derived from the calibration
equations. Overall, our photometric uncertainty is less than 10\%.

In order to properly subtract the continuum from the H$\alpha$ frames, we scaled
the counts of the continuum frames until (unsaturated) stars and elliptical galaxies 
reached an average H$\alpha +$ [N{\sc ii}] equivalent width of 0~\AA. 
The net H$\alpha +$ [N{\sc ii}] photometry of the selected galaxies was
performed using the QPHOT command of the APPHOT package in IRAF. Aperture
photometry was carried out, in both the ON-band and continuum frames, for each
galaxy and subtracted to get the net H$\alpha +$ [N{\sc ii}] fluxes. 

\section{Object Selection}

We made extensive use of the NASA Extragalactic Database (NED) to search
for known galaxies in the area covered by the observations. 
We measured the H$\alpha +$ [N{\sc ii}] fluxes for all
galaxies with known radial velocities, thus up to $r'
\approx 15.5$ for Abell~1367 and $r' \approx 16.5$ for Coma.

The narrow band filter used did not cover the whole
velocity interval of the clusters. In order to avoid large uncertainties in the
determination of fluxes and equivalent widths, we measured only galaxies for
which the filter transmitance was larger than 0.5.

   \begin{figure*}[!ht]
   \centering
      \caption{
Histograms of the H$\alpha$ fluxes (left) and equivalent widths (right) of the
emitting galaxies. Shaded bins correspond to the galaxies in Abell~1367.
}
         \label{ha_ew}
   \end{figure*}

Visual inspection of the net H$\alpha +$ [N{\sc ii}] frames allowed us to
identify faint galaxies with non negligible net H$\alpha +$ [N{\sc ii}]
emission. For these galaxies, there is no estimate of their velocities in
NED.
A population of faint galaxies ($r' \geq 17.2$) showed up, most of them
belonging to Abell~1367. Their H$\alpha +$ [N{\sc ii}] fluxes are low ($-15.53 < \log
F(\mbox{H}\alpha + \mbox{[N{\sc ii}]}) < -13.83$~erg~sec$^{-1}$~cm$^{-2}$) but
their H$\alpha$ equivalent width is in the range $9 <
EW(\mbox{H}\alpha + \mbox{[N{\sc ii}]}) < 418$\AA.
The search was performed for all objects visible on the NET-frames, but,   
in order to avoid spurious detections, we considered only objects with
$F(\mbox{H}\alpha + \mbox{[N{\sc ii}]})/\sigma_{flux} > 5$ as reliable detections
(see Column (9) of Tables ~\ref{ha_list_ab} and~\ref{ha_list_coma} for 
the definition of $\sigma_{flux}$). 
Since for some of them, their redshift is unknown, both the H$\alpha$ fluxes
and equivalent widths were computed assuming that their velocity coincides with
the average velocity of the cluster\footnote{the transmitance peak of
the filter is 0.85, and the transmitance for the H$\alpha$ lines at the
averages velocities of both clusters corresponds to 0.82 and 0.76 for Coma
and Abell~1367 respectively}$^{,}$\footnote{Contamination due the H$\beta$ or [O{\sc
iii}] emission lines of background objects is not important. For some of these
objects the redshift was measured and only less than 10\% were found to be non
members of the cluster}. 

In total 41 and 22 H$\alpha$ emitting galaxies were detected in
Abell~1367 and Coma respectively. These are listed in 
Tables~\ref{ha_list_ab}
and~\ref{ha_list_coma}, arranged as follows:

   \begin{figure*}
   \centering
      \caption{
Comparison between our fluxes (left) and equivalent widths (right)
and the values reported in the literature for some galaxies in common. Fluxes
are expressed in erg~sec$^{-1}$~cm$^{-2}$ and equivalent widths in \AA.
}
         \label{haflux_comp}
   \end{figure*}

\begin{itemize}
\item Col (1): Galaxy designation.
\item Col (2): CGCG name (Zwicky et al, 1961-1968).
\item Col (3): Other source designation.
\item Col (4, 5): Celestial coordinates (J2000).
\item Col (6): Radial velocity in km~sec$^{-1}$. Galaxies flagged with $\dagger$
 are those detected in H$\alpha$ but with unknown redshift. For these
objects, the mean velocity of the cluster was adopted. 
Galaxies flagged with $\dagger\dagger$ are those
for which a spectroscopic follow-up was carried out in
order to measure their redshifts (Gavazzi et al 2002, in prep). 
\item Col (7): Magnitude in the $r'$ band.
\item Col (8): Log of the H$\alpha +$[N{\sc ii}] flux, in erg~sec$^{-1}$~cm$^{-2}$.
\item Col (9): Error ($\sigma_{flux}$) of the H$\alpha +$[N{\sc ii}] flux.
The uncertainty includes three contributions: the Poisson photon counts error, 
the uncertainty on the background and the photometric uncertainty, which
is assumed as 10\% of the net flux. These errors were determined separately on
the ON and OFF-band frames, and combined using the standard error propagation.   
\item Col (10): H$\alpha +$[N{\sc ii}] equivalent width in \AA.
\item Col (11): Error in the H$\alpha +$[N{\sc ii}] equivalent width, computed similarly to
$\sigma_{flux}$ (Column (9)) except that the error on the absolute flux scale
does not affect the equivalent width.
\end{itemize}

Figure~\ref{ha_ew} shows the histograms of H$\alpha$ fluxes and equivalent
widths of the emitting galaxies. 

In order to check the quality of the photometry, we compared our fluxes and
equivalent widths with those taken from the literature (see Table~\ref{ha_comp}). 
Figure~\ref{haflux_comp} shows the plots of the H$\alpha$ fluxes
and equivalent widths reported in other works $vs$ ours. The linear regressions found
for both plots are the following:
\begin{displaymath}
\log F(\mbox{H}\alpha+\mbox{[N{\sc ii}]})_{this~work} = 
\end{displaymath}
\begin{equation}
0.34(\pm 0.91) + 1.03(\pm 0.07) \times \log
F(\mbox{H}\alpha+\mbox{[N{\sc ii}]})_{literature}
\end{equation}
\begin{displaymath}
\log EW(\mbox{H}\alpha+\mbox{[N{\sc ii}]})_{this~work} = 
\end{displaymath}
\begin{equation}
0.24(\pm 0.11) + 0.85(\pm 0.07) \times \log
EW(\mbox{H}\alpha+\mbox{[N{\sc ii}]})_{literature}
\end{equation}
Both plots show a discordant point, which corresponds with galaxy
CGCG~097-114. This galaxy was measured by Kennicutt et al. (1984), Moss et
al. (1988), Moss et al. (1998) and Gavazzi et al. (1998). There is agreement in
the flux between our data and that of Kennicutt et al. (1984) but not in the
equivalent width. The measurements by Moss et al. (1988, 1998) are
consistent with each other, but not with other sources. The equivalent
width by Gavazzi et al. (1998) is fairly consistent with ours.

\section{Discussion}

\subsection{The H$\alpha$ Luminosity Function}

The H$\alpha$ luminosity functions were computed separately for the two clusters under
study from the measured fluxes. 
$H_{0} = 50$~km~sec$^{-1}$~Mpc$^{-1}$ is assumed to allow a direct comparison
with Gallego et al. (1995).

H$\alpha$ fluxes were corrected for [N{\sc ii}] contamination and dust
extinction. 
The first correction is the one proposed by
Gavazzi et al. (2002, in prep.), based on the relationship found between the
$H$ band luminosities and the [N{\sc ii}]/H$\alpha$ ratio. 
After a empirical relationship between the $H$ and $r'$ magnitudes for the
galaxies in common in both samples the correction was finally given by:
\begin{equation}
\log [\mbox{N{\sc ii}}]/\mbox{H}\alpha = 1.26 - 0.19 \times r' + 0.70 \times \log D
\end{equation}
$D$ being the distance of the galaxies in Mpc.

The morphological type dependent dust extinction correction was taken from
Boselli et al. (2001). 
For galaxies with known morphological type (from NED or other sources), the correction was taken to be
\[
\begin{array}{ll}
A(\mbox{H}\alpha) = & \mbox{1.1 mag, for type Scd or earlier} \\
                    & \mbox{0.6 mag, for type Sd or later} \\
\end{array}
\]
For unclassified galaxies  we adopted $M_{B} = -18.25$ as the statistical
limiting magnitude for galaxy types intermediate between Scd and Sd, from
Sandage et al. (1985).

   \begin{figure*}
   \centering
      \caption{
Distributions of the radial velocities of the galaxies in Abell~1367 (upper
plot) and Coma (lower plot). The velocity distributions of the clusters assumed
gaussians are plotted with dashed lines. The shaded regions correspond to the
range of velocities not covered because of the low transmitance of the
filter. Only galaxies with known redshift were included in the plots.
}
         \label{velo_haflux}
   \end{figure*}

The contribution of active nuclei to the H$\alpha$ detections is negligible
because no relevant point-like nuclear features were 
detected in the H$\alpha$ frames.

In order to normalize the luminosity function to a proper volume,
angular radii of 3 and 4 degrees were assumed for Abell~1367 and Coma
respectively (Gavazzi et al. 1995), 
corresponding to linear sizes of 4.6 and 6.5~Mpc. The clusters were assumed 
spherically symmetric, thus the surveyed volume corresponds
to the intersection between the solid angle covered by our observations 
and the sphere containing the clusters.

A statistical correction was applied to account for the incomplete velocity coverage 
of the adopted [S{\sc ii}] filter.
Figure~\ref{velo_haflux} shows the flux distribution of galaxies
with known redshift $versus$ their radial velocities. The dashed line represents
the gaussian distributions of velocities described in Section~2. The shaded
regions correspond to the velocity ranges excluded from the filter transmitance window
for each cluster.
We estimate that about 20\% of the velocity distribution for Abell~1367 and 11\%
for Coma are not within the transmitance window of the narrow band filter.
We also corrected in a consistent way the effects of
the velocity distribution of the H$\alpha$ emitting galaxies with unknown
redshift. The correction was performed as follows:
first, we randomly distributed the velocities of these galaxies following the
gaussian probability density function with mean velocities and dispersions as
described in Section~2. New H$\alpha$ fluxes were derived for these galaxies,
according to the values of the transmitance of the [S{\sc ii}] filter,
for the randomly chosen velocities. If the assigned velocity of any
of these galaxies gave a transmitance $<$50\%, the object
was discarded. The final correction was performed by assuming that the
relationship, if any, between the radial velocities of the galaxies and the
H$\alpha$ fluxes should be symmetric with respect to the mean velocity of the
cluster.  
We repeated this procedure ten times in order to estimate the
statistical uncertainties induced by this effect on the luminosity function. Thus,
H$\alpha$ luminosity functions were computed with ten different flux
distributions for each cluster.

The functional form assumed for the LF is the Schechter (1976) function:
\begin{equation}
\phi(L)dL = \phi^{*}(L/L^{*})^{\alpha}\mbox{exp}(-L/L^{*})d(L/L^{*})
\end{equation}
The size of the bins was taken to be $\delta\log L = 1.0$ in order to minimize
the statistical errors. 
Table~\ref{counts} shows the  number counts per
luminosity interval for both clusters, averaged over the different random
distributions of velocities for the galaxies with unknown redshift. 

Table~\ref{fit} lists the obtained best fitting Schechter parameters
of the upper and lower envelopes for each cluster, as well as
the parameters for the average LFs finally adopted.

The upper and lower envelope H$\alpha$
LFs of the two clusters  are given in 
Figure~\ref{lf_field}.  Shaded regions between the
envelopes show the range of uncertainty of the H$\alpha$ LF for each cluster. 
The points correspond to the mean values listed in Table~\ref{counts},
and the error bars show their typical poissonian uncertainties. 
As reference, we plot the H$\alpha$ LFs of field galaxies obtained by Gallego et
al. (1995), Tresse \& Maddox (1998) and Sullivan et al. (2000). The 
lines are truncated at the completeness limits of each sample. 

   \begin{table}[b]
      \caption[]{Counts per luminosity bin}
         \label{counts}
     $$ 
         \begin{tabular}{lcc}
            \hline
            \noalign{\smallskip}
$\log L(\mbox{H}\alpha)$ & \multicolumn{2}{c}{Av. Number of gal.} \\
erg~sec$^{-1}$ & Abell~1367 & Coma \\
            \noalign{\smallskip}
            \hline
            \noalign{\smallskip}
38.8 & 8 & 2 \\
39.8 & 18 & 11 \\
40.8 & 13 & 8 \\
41.8 & 1 & 1 \\
            \noalign{\smallskip}
            \hline
         \end{tabular}
     $$ 
   \end{table}

   \begin{table}[b]
      \caption[]{Best fitting parameters for the upper and lower envelopes
corresponding to Abell~1367 and Coma. Also, the average adopted parameters for
the H$\alpha$ LF are listed.
}
         \label{fit}
     $$ 
         \begin{tabular}{lccc}
            \hline
            \noalign{\smallskip}
 & $\log \phi^{*}$ & $\alpha$ & $\log L^{*}$ \\
 & Mpc$^{-3}$ & & erg~sec$^{-1}$ \\
            \noalign{\smallskip}
            \hline
            \noalign{\smallskip}
\multicolumn{4}{c}{Abell~1367} \\
Upper envelope & $-0.06$ & $-0.94$ & 41.37 \\
Lower envelope & $+0.20$ & $-0.72$ & 41.21 \\
Average & $+0.06$ & $-0.82$ & 41.30 \\
\hline
\multicolumn{4}{c}{Coma} \\
Upper envelope & $-0.09$ & $-0.70$ & 41.24 \\
Lower envelope & $-0.04$ & $-0.53$ & 41.21 \\
Average & $-0.07$ & $-0.60$ & 41.23 \\
            \noalign{\smallskip}
            \hline
         \end{tabular}
     $$ 
   \end{table}

Disregarding non-completeness effects, which should
only affect our lowest luminosity bins,
the LFs of the two clusters are in fair agreement.
The apparent difference with the field LFs is mainly in the normalization
since the density of galaxies is several orders of magnitude larger
in clusters than in the field.
Beside the normalization, the shape of the cluster LFs appears steeper at the bright end and flatter
at the faint end.  
The former derives from undersampling at high luminosity (due to small volume coverage
in the two clusters we do not detect any object with $F(H\alpha) \geq
10^{42}$~erg~sec$^{-1}$ as opposed to Gallego et al. 1996). 

The slope of the fitted LFs appear different among clusters and field at the faint end.
However the data points, within the completeness limits of each survey, appear
in full agreement among each other, as shown in Fig. \ref{pru_lf}.

   \begin{figure}[!t]
   \centering
      \caption{
H$\alpha$ luminosity functions for Abell~1367 (diamonds) and for Coma
(asterisks). The best fittings to Schechter functions are shown together
with those found for the local Universe (Gallego et al. 1995), for $z
\approx 0.2$ (Tresse et al. 1998) and for a sample of UV selected galaxies
(Sullivan et al. 2000). 
The shaded regions show the uncertainty region between the lower and upper
envelopes. 
}
         \label{lf_field}
   \end{figure}

In this Figure we scaled the cluster LFs in such a  way that they match 
the field LF at $\log L(\mbox{H}\alpha) \approx 41$~erg~sec$^{-1}$. 
Above $\log L(\mbox{H}\alpha)\approx 40$~erg~sec$^{-1}$, where all the samples are complete,
there is consistency between the field and the cluster datasets. Nothing can be
said for fainter  luminosities because the field samples are incomplete or
present rather poor statistics, opposite to the present cluster survey which is
complete to $\log L(\mbox{H}\alpha) \approx 39$~erg~sec$^{-1}$. Deeper H$\alpha$
surveys of the field are necessary to assess if the  differences at the faint
luminosity end are significant.

\subsection{The Virgo Cluster}

It is instructive to compare the H$\alpha$ LF of A1367 and Coma with that of the
Virgo cluster. Given its large angular size, performing  
a complete H$\alpha$ survey of this cluster would be prohibitive. 
However H$\alpha$ observations
of most of the brightest galaxies (230 objects brighter than B=16 mag) are available 
(Boselli \& Gavazzi 2002; Gavazzi et al. 2002). 
Using these data we construct a "pseudo" H$\alpha$ LF by transforming the
B band LF into an H$\alpha$ one after having shown that H$\alpha$ luminosity
and $M_B$ are found proportional one-another.

Figure~\ref{virgo_rel} shows the H$\alpha$ luminosity $vs$ the absolute $M_B$
magnitude relationship. Distances are estimated according to the Virgo cluster
group membership, as defined in Gavazzi \& Boselli (1999). 
The best fit to the data gives a slope of 0.37, consistent with 0.40 (i.e. a slope
of 1 in a luminosity-luminosity plot). For simplicity we adopted this last value, 
because it allows to transform the observed B band Schechter function into an H$\alpha$ LF
of the same functional form.
Therefore we adopt: 
\begin{equation}
\label{havb}
\log L(\mbox{H}\alpha) = -0.40 \times M_{B} + 33.12
\end{equation}
Combining this relationship with the $B$ band
luminosity function for spirals and irregulars in the Virgo
core obtained by Sandage et al. (1985) we obtain an H$\alpha$ LF:
\begin{equation}
\phi'(L) = 1.07 \times (L/10^{41.2})^{-0.8}~\mbox{exp}[-(L/10^{41.2})]
\end{equation}

Figure~\ref{lf_clusters} shows the Virgo H$\alpha$ LF together with the ones
obtained for Abell~1367 and Coma. The shaded region reflects the 
scatter in the
relationship between H$\alpha$ luminosities and $M_B$ found in Virgo.
The shape of the Virgo LF appears consistent with that of Abell~1367
and Coma, despite the different nature of the three clusters, Virgo being
unrelaxed and spiral rich, Abell~1367 relaxed and spiral rich, and Coma relaxed
and spiral poor. 

\subsection{Star Formation Rates in Clusters}

The total star formation rate per unit volume for
clusters is derived by integrating the best fitting Schechter functions over
the whole range of luminosities. 
To be consistent with Gallego et al. (1995), we convert the H$\alpha$
luminosities to star formation per unit time using:
\begin{equation}
L(\mbox{H}\alpha) = 9.40 \times 10^{40}
\frac{\mbox{SFR}}{M_{\odot}~\mbox{yr}^{-1}}\mbox{erg~sec}^{-1}
\end{equation}
Total integrated SFRs of 2.20 and 1.36~$M_{\odot}$~yr$^{-1}$~Mpc$^{-3}$ are obtained for Abell~1367
and Coma respectively, i.e.
more than two orders of magnitude higher than the value of 0.013 $M_{\odot}$~yr$^{-1}$~Mpc$^{-3}$ 
reported for the local Universe by Gallego et al. (1995). 

   \begin{figure}[t]
   \centering
      \caption{
Galaxy number density per unit volume $vs$ the H$\alpha$ luminosity for the
clusters and for the different field samples. The cluster counts have been
normalized to properly match the field counts. 
}
         \label{pru_lf}
   \end{figure}

The estimate of the contribution of the clusters to the
total SFR per unit volume of the local Universe, is obtained by taking into account
the local spatial density of clusters. For Abell type 2 clusters, like
Abell~1367 and Coma, this value was reported to be $1.84 \times
10^{-5}$~Mpc$^{-3}$ (Bramel et al. 2000), although this number is affected by
large uncertainties. We conclude that the typical contribution of Abell type 2
clusters to the SFR per unit volume is about $3.3 \times
10^{-5}$~$M_{\odot}$~yr$^{-1}$, that is 0.25\% of the total SFR in the local
Universe. 

Similarly, by integrating the Virgo H$\alpha$ luminosity function, we obtain a total
H$\alpha$ luminosity density of $1.56 \times
10^{41}$~erg~sec$^{-1}$~Mpc$^{-3}$, which gives a SFR of
1.65~$M_{\odot}$~yr$^{-1}$~Mpc$^{-3}$. Taking into account that the
Virgo cluster is classified as Abell type 1 (Struble \& Rood 1982), and
assuming the spatial density for clusters of this type (Bramel et
al. 2000) of $8.46 \times 10^{-4}$~Mpc$^{-3}$, we obtain that the contribution of
type 1 clusters is $1.40 \times
10^{-3}$~$M_{\odot}$~yr$^{-1}$~Mpc$^{-3}$, corresponding to 10.8\% of the total SFR
density in the local Universe.

\section{Conclusions}

We have carried out an H$\alpha$ imaging survey of the central 1~deg$^{2}$ of the
nearby clusters Abell~1367 and Coma. Significant H$\alpha$ emission is found
associated with 41 galaxies in Abell~1367 and 22 in Coma. These data
are used to estimate, for the first time, the H$\alpha$ luminosity function of 2 nearby
clusters of galaxies. These LFs are found consistent with the H$\alpha$ luminosity function
derived for the Virgo cluster, despite their different nature.
The typical Schechter parameters: $\phi^{*} \approx 10^{0.00\pm0.07}$~Mpc$^{-3}$, $L^{*} \approx
10^{41.25\pm0.05}$~erg~sec$^{-1}$ and $\alpha \approx -0.70\pm0.10$ are obtained. 

   \begin{figure}
   \centering
      \caption{
$M_{B}$~$vs$~$\log L(\mbox{H}\alpha)$ for a large sample of Virgo galaxies. The
solid line indicates the adopted fit, the dashed line shows the best fit
obtained from a least squares fitting. A distance modulus of 31.7~mag was
adopted to convert observed fluxes and magnitudes to luminosities and absolute
magnitudes. 
}
         \label{virgo_rel}
   \end{figure}

The best fitting parameters of the cluster LFs are significantly different from
those found for field galaxies, in particular at the faint end where
the cluster slope is shallower than the extrapolated slope of the field LF.
However it must be stressed that the steep slope found in the field
is based on relatively high luminosity points and no data are available
below  $\log L(\mbox{H}\alpha) \approx 40$~erg~sec$^{-1}$ i.e. 
where the cluster LFs begin to flatten out. 
After re-normalizing the cluster data on the field ones, the two sets of data points 
are found consistent within the completeness limit of the field samples. 
Until a deeper field LF will be available it is impossible to
establish whether the apparent underabundance of low luminosity objects in clusters
is a real evolutionary effect or it is an artifact due to incompleteness.

By computing the total SFR per unit volume of the cluster galaxies, and taking into
account the cluster density in the local Universe, we estimate that the
contribution of type 2 and type 1 clusters is  about 0.25\% and 10.8\% respectively
of the SFR per unit volume of the local Universe.

   \begin{figure}
   \centering
      \caption{
Same as Figure~6.
We included the expected curve for the
Virgo cluster assuming the $B$ band luminosity function from Sandage et
al. (1985) and the $L(\mbox{H}\alpha)~vs~M_{B}$ relationship given by equation~5.
}
         \label{lf_clusters}
   \end{figure}

\begin{acknowledgements}
This research has made use of the NASA/IPAC Extragalactic Database (NED) which
is operated by the Jet Propulsion Laboratory, California Institute of
Technology, under contract with the National Aeronautics and Space
Administration. The INT is operated on the island of La Palma 
by the ING group, in the Spanish Observatorio del 
Roque de Los Muchachos of the Instituto de Astrof\'{\i}sica 
de Canarias. 
\end{acknowledgements}

\onecolumn
   \begin{table}[t]
      \caption[]{Some properties of the selected H$\alpha$ emitting galaxies in Abell~1367}
         \label{ha_list_ab}
     $$ 
         \begin{tabular}{lccccccccrr}
            \hline
            \noalign{\smallskip}
Name & CGCG & Other & R.A. & Dec. & $v_{r}$ & $r'$ & $F_{\alpha}$ &
$\Delta_{f}$ & 
W$_{\alpha}$ & $\Delta_{\scriptsize{\mbox{W}}}$  \\
            \noalign{\smallskip}
            \hline
            \noalign{\smallskip}
114024+195747 & --- & --- & 11 40 24.90 & +19 57 47.7 & 6749            &
15.48 & $-$13.71 & 0.04 & 14 & 1 \\
114038+195437 & --- & --- & 11 40 38.96 & +19 54 37.4 & 6500$\dagger\dagger$        &
17.35 & $-$14.09 & 0.05 & 36 & 3 \\
114107+200251 & --- & --- & 11 41 07.79 & +20 02 51.3 & 6500$\dagger$   &
18.91 & $-$14.60 & 0.05 & 43 & 4  \\
114110+201117 & --- & --- & 11 41 10.47 & +20 11 17.7 & 6500$\dagger$   &
17.57 & $-$13.95 & 0.04 & 56 & 2  \\
114112+200109 & --- & --- & 11 41 12.81 & +20 01 09.9 & 6500$\dagger$   &
19.44 & $-$14.80 & 0.07 & 38 & 5  \\
114141+200230 & --- & --- & 11 41 41.20 & +20 02 30.5 & 6500$\dagger$   &
17.37 & $-$14.26 & 0.06 & 26 & 4  \\
114142+200054 & --- & --- & 11 41 42.57 & +20 00 54.9 & 6500$\dagger$   &
17.33 & $-$14.36 & 0.07 & 19 & 3 \\
114149+194605 & --- & --- & 11 41 49.79 & +19 46 05.1 & 6500$\dagger$   &
17.52 & $-$14.37 & 0.05 & 23 & 2  \\
114156+194207 & --- & --- & 11 41 56.69 & +19 42 07.8 & 6500$\dagger$   &
19.77 & $-$15.10 & 0.07 & 36 & 5  \\
114157+194329 & --- & --- & 11 41 57.90 & +19 43 29.4 & 6500$\dagger$   &
20.21 & $-$15.29 & 0.05 & 34 & 3  \\
114158+194149 & --- & --- & 11 41 58.05 & +19 41 49.6 & 6500$\dagger$   &
19.46 & $-$15.02 & 0.06 & 38 & 4  \\
114158+194205 & --- & --- & 11 41 58.10 & +19 42 05.9 & 6500$\dagger$   &
20.30 & $-$15.27 & 0.04 & 49 & 2   \\
114158+194900 & --- & --- & 11 41 58.26 & +19 49 00.9 & 6500$\dagger$   &
20.70 & $-$15.53 & 0.07 & 32 & 4  \\
114202+194348 & --- & --- & 11 42 02.30 & +19 43 48.5 & 6500$\dagger$   &
20.83 & $-$15.35 & 0.05 & 70 & 6  \\
114202+192648 & --- & --- & 11 42 02.96 & +19 26 48.2 & 6500$\dagger$   &
19.54 & $-$14.66 & 0.06 & 32 & 4  \\
114214+195833 & 097-062 & PGC036330 & 11 42 14.55 & +19 58 33.6 & 7815 &
14.51 & $-$13.19 & 0.04 & 28 & 1 \\
114215+200255 & 097-063 & PGC036323 & 11 42 15.70 & +20 02 55.2 & 6102 &
15.36 & $-$13.69 & 0.04 & 13 & 1 \\
114218+195016 & --- & --- & 11 42 18.08 & +19 50 16.1 & 6476            &
15.79 & $-$14.24 & 0.04 & 6 & 1 \\
114239+195808 & --- & --- & 11 42 39.23 & +19 58 08.0 & 7345            &
16.95 & $-$13.89 & 0.04 & 40 & 1 \\
114240+195716 & --- & --- & 11 42 40.36 & +19 57 16.6 & 6500$\dagger$   &
17.68 & $-$14.70 & 0.08 & 13 & 2  \\
114256+195757 & 097-073 & PGC036382 & 11 42 56.67 & +19 57 57.7 & 7275 &
15.50 & $-$12.81 & 0.04 & 86 & 1  \\
114313+193645 & --- & --- & 11 43 13.08 & +19 36 45.8 & 6500$\dagger\dagger$            &
17.27 & $-$14.06 & 0.05 & 30 & 3  \\
114313+200015 & 097-079 & PGC036406 & 11 43 13.93 & +20 00 15.6 & 7000 &
16.50 & $-$12.69 & 0.04 & 130 & 2  \\
114341+200135 & --- & --- & 11 43 41.62 & +20 01 35.3 & 6500$\dagger$   &
17.08 & $-$14.15 & 0.06 & 25 & 3  \\
114348+195812 & 097-087 & UGC06697 & 11 43 48.59 & +19 58 12.8 & 6725  &
14.22 & $-$12.19 & 0.04 & 81 & 2  \\
114348+201456 & --- & --- & 11 43 48.92 & +20 14 56.0 & 6146            &
15.86 & $-$12.95 & 0.04 & 137 & 1  \\
114349+195833 & --- & --- & 11 43 49.87 & +19 58 33.2 & 7542            &
16.11 & $-$13.99 & 0.04 & 19 & 2  \\
114355+192743 & --- & --- & 11 43 55.71 & +19 27 43.9 & 6500$\dagger$   &
18.72 & $-$14.67 & 0.07 & 27 & 4  \\
114358+201105 & 097-092 & PGC036478 & 11 43 58.17 & +20 11 05.6 & 6373 &
14.71 & $-$13.10 & 0.04 & 30 & 1  \\
114358+200433 & 097-091 & NGC3840 & 11 43 58.81 & +20 04 33.0 & 7368   &
13.92 & $-$12.86 & 0.07 & 25 & 4  \\
114400+200144 & 097-097 & NGC3844 & 11 44 00.86 & +20 01 44.5 & 6834   &
13.62 & $-$13.41 & 0.04 & 5 & 1  \\
114430+195718 & --- & --- & 11 44 30.41 & +19 57 18.8 & 6500$\dagger$   &
20.23 & $-$14.38 & 0.04 & 418 & 19 \\
114447+194624 & 097-114 & NGC3860B & 11 44 47.88 & +19 46 24.6 & 8293  &
15.33 & $-$13.24 & 0.05 & 40 & 4  \\
114454+194733 & --- & --- & 11 44 54.22 & +19 47 33.2 & 6500$\dagger\dagger$            &
20.27 & $-$13.99 & 0.06 & 103 & 17  \\
114454+194635 & 097-125 & PGC036589 & 11 44 54.99 & +19 46 35.8 & 8271 &
14.50 & $-$13.00 & 0.05 & 24 & 2  \\
114454+200101 & --- & --- & 11 44 54.71 & +20 01 01.5 & 6500$\dagger\dagger$            &
16.17 & $-$14.41 & 0.04 & 6 & 1  \\
114503+195002 & --- & --- & 11 45 03.38 & +19 50 02.7 & 6500$\dagger$   &
17.90 & $-$14.76 & 0.07 & 9 & 1  \\
114506+195801 & 097-129E & NGC3861B & 11 45 06.91 & +19 58 01.6 & 6009 &
14.64 & $-$13.38 & 0.06 & 19 & 2  \\
114513+194523 & --- & --- & 11 45 13.86 & +19 45 23.0 & 6500$\dagger\dagger$            &
15.60 & $-$13.86 & 0.04 & 12 & 1  \\
114518+200009 & --- & --- & 11 45 18.00 & +20 00 09.5 & 6500$\dagger$   &
17.54 & $-$14.28 & 0.06 & 22 & 3  \\
114603+194712 & 097-143B & --- & 11 46 03.68 & +19 47 12.9 & 7170       &
15.80 & $-$14.93 & 0.05 & 1 & 1  \\
            \noalign{\smallskip}
            \hline
         \end{tabular}
     $$ 
\begin{list}{}{}
\item[$\dagger$] Objects with unknown redshift but detected in the net H$\alpha$
frames
\item[$\dagger\dagger$] Objects for which we have measured the redshift. It will
appear in a subsequent paper.
\end{list}
   \end{table}

   \begin{table}[t]
      \caption[]{Some properties of the selected H$\alpha$ emitting galaxies in Coma}
         \label{ha_list_coma}
     $$ 
         \begin{tabular}{lccccccccrr}
            \hline
            \noalign{\smallskip}
Name & CGCG & Other & R.A. & Dec. & $v_{r}$ & $r'$ & $F_{\alpha}$ &
$\Delta_{f}$ & 
W$_{\alpha}$ & $\Delta_{\scriptsize{\mbox{W}}}$  \\
            \noalign{\smallskip}
            \hline
            \noalign{\smallskip}
125757+280343 & --- & FOCA610 & 12 57 57.73 & +28 03 43.3 & 8299 & 15.23 &
$-$13.37 & 0.05 & 22 & 2 \\
125805+281433 & 160-055 & NGC4848 & 12 58 05.67 & +28 14 33.2 & 7049 &
14.04 & $-$12.54 & 0.05 & 34 & 2 \\
125845+284133 & --- & FOCA353 & 12 58 45.64 & +28 41 33.1 & 7001$\dagger\dagger$
& 17.21 & $-$14.02 & 0.06 & 35 & 5 \\
125845+283235 & --- & FOCA399 & 12 58 45.80 & +28 32 35.3 & 7001$\dagger$
& 17.76 & $-$13.83 & 0.04 & 101 & 4  \\
125856+275002 & 160-212 & FOCA600 & 12 58 56.55 & +27 50 2.7  & 7378 &
15.12 & $-$13.84 & 0.05 & 3 & 1  \\
125902+280656 & 160-213 & FOCA498 & 12 59 02.14 & +28 06 56.4 & 9436 &
15.15 & $-$13.32 & 0.06 & 28 & 3 \\
125907+275118 & 160-219 & IC3960 & 12 59 07.97 & +27 51 18.0 & 6650 &
14.50 & $-$14.12 & 0.05 & 2 & 1  \\
125923+282919 & --- & FOCA361 & 12 59 23.13 & +28 29 19.0 & 7001$\dagger\dagger$
& 15.75 & $-$13.98 & 0.04 & 10 & 1  \\
130006+281500 & --- & FOCA371 & 13 00 06.42 & +28 15 0.9  & 7259 & 17.04 & 
$-$14.48 & 0.07 & 6 &  1 \\
130037+280327 & 160-252 & FOCA388 & 13 00 37.99 & +28 03 27.6 & 7840 &
14.68 & $-$12.93 & 0.08 & 41 & 4 \\
130037+283951 & --- & --- & 13 00 37.24 & +28 39 51.6 & 7001$\dagger\dagger$ &
16.86 & $-$14.64 & 0.05 & 6 &  2 \\
130040+283113 & --- & FOCA242 & 13 00 40.75 & +28 31 13.4 & 8901 & 15.80 & 
$-$13.11 & 0.05 & 68 & 6 \\
130056+274727 & 160-260 & FOCA445 & 13 00 56.03 & +27 47 27.7 & 7985 &
13.11 & $-$12.76 & 0.07 & 11 & 2 \\
130114+283118 & --- & FOCA195 & 13 01 14.99 & +28 31 18.5 & 8426 & 17.02 & 
$-$14.04 & 0.05 & 29 & 3 \\
130125+284036 & 160-098 & FOCA137 & 13 01 25.04 & +28 40 36.9 & 8762 &
14.41 & $-$13.21 & 0.04 & 18 & 1 \\
130127+275957 & --- & GMP2048 & 13 01 27.17 & +27 59 57.0 & 7558 & 15.64 & 
$-$14.35 & 0.04 & 4 & 1  \\
130128+281515 & --- & --- & 13 01 28.63 & +28 15 15.9 & 7001$\dagger$ &
20.41 & $-$14.96 & 0.04 & 107 & 6  \\
130130+283328 & --- & FOCA158 & 13 01 30.85 & +28 33 28.0 & 7001$\dagger\dagger$ & 16.76 &
$-$13.95 & 0.06 & 24 & 2 \\
130140+281456 & --- & GMP1925 & 13 01 40.97 & +28 14 56.6 & 7001$\dagger$
& 19.33 & $-$14.43 & 0.07 & 132 & 36  \\
130158+282114 & --- & --- & 13 01 58.43 & +28 21 14.8 & 7001$\dagger$ &
19.81 & $-$14.39 & 0.04 & 278 & 8 \\
130212+281023 & --- & FOCA218 & 13 02 12.00 & +28 10 23.0 & 8950 & 16.09 &
$-$13.41 & 0.05 & 30 & 2 \\
130212+281253 & 160-108 & FOCA204 & 13 02 12.55 & +28 12 53.0 & 8177 &
14.93 & $-$13.29 & 0.04 & 25 & 1 \\
            \noalign{\smallskip}
            \hline
         \end{tabular}
     $$ 
   \end{table}
\twocolumn

\newpage

\onecolumn
   \begin{table}[t]
      \caption[]{Comparison of the H$\alpha$ fluxes and equivalent widths with
data from the literature, for the objects in common.}
         \label{ha_comp}
     $$ 
         \begin{tabular}{lcccccccccccc}
            \hline
            \noalign{\smallskip}
CGCG & \multicolumn{2}{c}{This work} & 
\multicolumn{2}{c}{M88$^{\mathrm{a}}$} & 
\multicolumn{2}{c}{M98$^{\mathrm{b}}$} & 
\multicolumn{2}{c}{K84$^{\mathrm{c}}$} & 
\multicolumn{2}{c}{G91$^{\mathrm{d}}$} & 
\multicolumn{2}{c}{G98$^{\mathrm{e}}$} \\
 & $\log F$ & $EW$ & 
$\log F$ & $EW$ & 
$\log F$ & $EW$ & 
$\log F$ & $EW$ & 
$\log F$ & $EW$ & 
$\log F$ & $EW$ \\
            \noalign{\smallskip}
            \hline
            \noalign{\smallskip}
097-062 & -13.19 &  28 & -- &  -- & -12.93 &  58 & -13.10 &  45 & -- & -- & -- &  34 \\
097-073 & -12.81 &  86 & -12.84 &  -- & --- &  --- & -12.84 &  80 & -- & -- & -12.76 & 108 \\
 &  &  &  &  &  &  &  &  &  &  & -12.75 &  94 \\
097-079 & -12.69 & 130 & -12.54 & --- & --- & --- & -12.64 & 145 & -12.64 & 131 & -12.66 & 137 \\
097-087 & -12.19 &  81 & -12.22 & 64 & -12.43 & 84 & -12.19 &  61 & -- & -- & -12.22 & 74 \\
097-092 & -13.10 &  30 & -13.06 & -- & -12.95 &  30 & -- & -- & -- & -- & -- &  27 \\
097-091 & -12.86 &  25 & -12.92 & 17 & -12.86 &  21 & -12.74 &  23 & -- & -- & -- & -- \\
097-114 & -13.24 &  40 & -12.82 & 79: & -12.82 & 60 & -13.20 &   4 & -- & -- & -- & 48 \\
097-125 & -13.00 &  24 & -13.13 &  29 & -13.04 & 26 & -- & -- & -- & -- & -- & 21 \\
097-129E & -13.38 &  19 & -- & -- & -- & -- & -- & -- & -- & -- & -13.38 & 18 \\
160-252 & -12.93 &  41 & --    & -- & --    & -- & -12.93 &  35 & -- & -- & -- & -- \\
160-055 & -12.54 &  34 & --    & -- & --    & -- & -12.65 &  23 & -- & -- & -12.51 & 34 \\
160-260 & -12.76 &  11 & -- & -- & -- & -- & -- & -- & -- & -- & -13.03 & 8 \\
160-098 & -13.21 &  18 & -- & -- & -- & -- & -- & -- & -- & -- & -13.15 & 20 \\
            \noalign{\smallskip}
            \hline
         \end{tabular}
     $$ 
\begin{list}{}{}
\item[$^{\mathrm{a}}$] Moss et al. 1988
\item[$^{\mathrm{b}}$] Moss et al. 1998
\item[$^{\mathrm{c}}$] Kennicutt et al. 1984
\item[$^{\mathrm{d}}$] Gavazzi et al. 1991
\item[$^{\mathrm{e}}$] Gavazzi et al. 1998
\end{list}
   \end{table}
\twocolumn

\newpage

\appendix

\section{Comments on Individual Objects}

The field around galaxies CGCG~97-114, CGCG~97-120 and CGCG~97-125 deserves
special attention. Each one of them shows H$\alpha$ emission. However 
CGCG~97-120 is not listed in Table~\ref{ha_list_ab} because its radial
velocity, 5595~km~sec$^{-1}$, (de Vaucouleurs, 1991) is out of the
transmitance window of the
on-band filter. Instead, CGCG~97-114 and CGCG~97-125 have very similar
velocities lying in our filter range: 8293~km~sec$^{-1}$ (de Vaucouleurs, 1991)
and 8271~km~sec$^{-1}$ (Haynes et al, 1997)
respectively. If these galaxies form a sub-group within Abell~1367
($<V> = 6500$~km~sec$^{-1}$), this entity must have  a very high velocity
dispersion. The group coincides in position with the X-ray cluster center
(Donnelly et al., 1998). CGCG~97-120  is a very H{\sc i} deficient galaxy
(def(H{\sc i}) = 0.9,
Chincarini et al., 1983), indicating that it has passed through the cluster
center,  while CGCG~97-125 is not (def(H{\sc i}) = $-0.21$, Haynes et al.,
1997). No H{\sc i}
measurement is available for CGCG~97-114,because it lies close (in space and in
velocity) to CGCG~97-125.

In between CGCG~97-120 and CGCG~97-125 we found two emission-line dwarf galaxies
114454+194733 and 114451+194713  (labeled as F.O. but not included in our list
because its S/N ratio is lower than 5) (see Figure~\ref{fo18}) which  were already
revealed by Sakai et al. (2001). For both galaxies a spectroscopic follow-up gave
velocities about 8100~km~sec$^{-1}$, consistent with those of CGCG~97-114 and CGCG~97-125,
indicating that they are probably associated with these galaxies.

In addition, our net frame 
shows the presence of some H$\alpha$ emission bridging CGCG~97-114 and 
CGCG~97-125. Two bright H$\alpha$ knots are detected near CGCG~97-125 followed by
a filamentary structure  near the bright star between the two galaxies.
This structure probably indicates a tidal interaction between the two galaxies,
consistently with the shells of ~1' diameter seen in the off-band image around
CGCG~97-125. In fact N-body simulations (Quinn, 1984; Seguin, 1996) confirm
Schweizer (1980) suggestion that shells around early type galaxies result from
galaxy collisions. It is not yet possible to decide whether  114454+194733 and
F.O. are tidal dwarfs of CGCG~97-125 or patches of star-forming gas stripped
from it.

\newpage

\onecolumn

   \begin{figure}[t]
   \centering
      \caption{
Net H$\alpha$ images of the selected galaxies. Blue contours correspond to the
continuum image.
}
         \label{images}
   \end{figure}

\addtocounter{figure}{-1}

\newpage

   \begin{figure}[t]
   \centering
      \caption{
Continued.
}
         \label{images}
   \end{figure}

\addtocounter{figure}{-1}

\newpage

   \begin{figure}[t]
   \centering
      \caption{
Continued.
}
         \label{images}
   \end{figure}

\addtocounter{figure}{-1}

\newpage

   \begin{figure}[t]
   \centering
      \caption{
Continued.
}
         \label{images}
   \end{figure}

\addtocounter{figure}{-1}

\newpage

   \begin{figure}[t]
   \centering
      \caption{
Continued.
}
         \label{images}
   \end{figure}

\newpage

   \begin{figure}[t]
   \centering
      \caption{
Detailed net H$\alpha$, on-band and off-band images of F.O. 18.
}
         \label{fo18}
   \end{figure}

\end{document}